\documentclass[letter]{jpsj2} 
\usepackage{bm}

\title{Spin Damping Monopole}

\author{Akihito \textsc{Takeuchi}\thanks{E-mail: atake@phys.se.tmu.ac.jp} and Gen \textsc{Tatara}}

\inst{
Department of Physics, Tokyo Metropolitan University, Hachioji, Tokyo 192-0397, Japan
}

\abst{
We present theoretical evidence that a magnetic monopole emerges in dynamic magnetic systems in the presence of the spin-orbit interaction.
The monopole field is expressed in terms of spin damping associated with magnetization dynamics.
We demonstrate that the observation of this spin damping monopole is accomplished electrically using Amp\`ere's law for monopole current.
Our discovery suggests the integration of monopoles into electronics, namely, monopolotronics.
}

\kword{
magnetic monopole,
spin damping,
inverse spin Hall effect,
Maxwell's equations
}

\begin{document}

\maketitle


The magnetic monopole predicted by Dirac in 1931~\cite{Dirac31} is a unique particle that arises from singularity~\cite{Jackson98}.
In high-energy physics, a monopole emerges if the electromagnetic interaction in the world, described by a U(1) algebra, is a result of the symmetry breaking of a unified force having a higher symmetry of SU(5)~\cite{tHooft74,Polyakov74}.
Intensive effort has been exerted to find a grand unified theory (GUT) monopole by waiting for a monopole created in the early universe to go through superconducting detectors~\cite{Cabrera82} and by detecting its ionization~\cite{macro02}; however, no evidence has been found thus far.
The energy needed to create a GUT monopole is about $10^{17}$ GeV, and so creating one in an accelerator on earth is impossible.

Since a monopole is a consequence of symmetry breaking, it exists in solids too.
Symmetry breaking in solids occurs at an energy much lower than the GUT energy, typically below 1 eV, and therefore experiments on this are feasible.
The most well-known monopole in solids is the hedgehog monopole (HHM), which arises in magnetic materials~\cite{Volovik87}; it is the SU(2) counterpart of the GUT monopole.
The key interaction for the HHM is the coupling of the conduction electron to the magnetization represented by the vector ${\bm M}({\bm r},t)$, which depends on the space coordinate ${\bm r}$ and the time $t$.
The electronic spin, represented by a vector ${\bm \sigma}$, is polarized by ${\bm M}$ owing to the coupling
\begin{equation}
H_{\rm sd}
=
-J {\bm M} \cdot {\bm \sigma},
\end{equation}
where $J$ is the coupling constant.
Since ${\bm M}$ is an external field for electrons, the SU(2) symmetry of the electronic spin is broken.
By diagonalizing $H_{\rm sd}$ and choosing the spin quantization direction to be along the $z$-axis, the electrons are described as spin-polarized and interacting with a SU(2) gauge field~\cite{Volovik87}, $A_\mu^a$ ($\mu = t,x,y,z$ and $a = x,y,z$ are the indices in the coordinate space and spin space, respectively).
In the adiabatic limit, i.e., when $J$ is large, only the $z$ component of the gauge field, $A_\mu^z$, survives and it acts as a U(1) gauge field.
The HHM originates from the deviation from the perfect U(1) symmetry, i.e., from the perpendicular components $A_\mu^x$ and $A_\mu^y$.
In fact, the electrons feel the SU(2) gauge fields whose strength is
$F_{\mu\nu}^a = \partial_\mu A_\nu^a -\partial_\nu A_\mu^a -(2e/\hbar) \epsilon_{abc} A_\mu^b A_\nu^c$,
where $e$ is the electric charge, $\hbar$ is the Planck constant divided by $2\pi$, and $\epsilon_{abc}$ represents the antisymmetric tensor.
When this field is projected onto the U(1) space, we obtain the electromagnetic field tensor as 
$F_{\mu\nu}^z = \partial_\mu A_\nu^z -\partial_\nu A_\mu^z +\Phi_{\mu\nu}$,
where $\Phi_{\mu\nu} \equiv -(2e/\hbar) (A_\mu^x A_\nu^y -A_\mu^y A_\nu^x)$.
The anomalous field strength $\Phi_{\mu\nu}$, which in terms of ${\bm n} \equiv  {\bm M} / {|M|}$ reads 
$\Phi_{\mu\nu} = -(\hbar/2e) {\bm n} \cdot (\partial_\mu {\bm n} \times \partial_\nu {\bm n})$,
represents the HHM. 
In fact, the field strength satisfies  
$(1/2) \epsilon_{\mu\nu\sigma\rho} \partial_\nu F_{\sigma\rho}^z = 0$, whose components read
${\bm \nabla} \times {\bm E} +\dot{\bm B} = -{\bm j}_{\rm h}$,
and 
${\bm \nabla} \cdot {\bm B} = \rho_{\rm h}$,
where the electric and magnetic fields are 
$E_i
\equiv -\nabla_i A^z_t -\dot{A}^z_i
= -(\hbar/2e) {\bm n} \cdot (\dot{\bm n} \times \nabla_i {\bm n})$
and
$B_i
\equiv \epsilon_{ijk} \nabla_j A^z_k
= (\hbar/4e) \epsilon_{ijk} {\bm n} \cdot (\nabla_j {\bm n} \times \nabla_k {\bm n})$,
respectively.
The monopole current ($j_{\rm h}$) and its density ($\rho_{\rm h}$) for the HHM are given as
$j_{{\rm h}, i} = -(3\hbar/4e) \epsilon_{ijk} \dot{\bm n} \cdot (\nabla_j {\bm n} \times \nabla_k {\bm n})$
and
$\rho_{\rm h} = (\hbar/4e) \epsilon_{ijk} \nabla_i {\bm n} \cdot (\nabla_j {\bm n} \times \nabla_k {\bm n})$,
respectively.
Although the HHM is mathematically allowed, experimental realization has not been achieved thus far.
In fact, as seen from the expression for $j_{{\rm h}}$ and  $\rho_{\rm h} $, the HHM disappears in common magnets where the length of the local magnetization, ${\bm n}$, is constant.
In addition, the boundary condition at infinity for the HHM would not be easy to realize experimentally.

In this paper, we search for a different monopole in magnets, which exists in conventional ferromagnets where the local magnetization length is constant.
Such a monopole current creates the rotational electric field via Amp\`ere's law.
This means that the monopole current is an anomalous angular momentum source that induces the rotational motion of electric charges.
To realize such a monopole in magnetic systems, a coupling between spin and electron orbital motion is, therefore, essential.
Such a coupling is known to emerge from a relativistic effect, namely, the spin-orbit interaction.
The spin-orbit interaction exists in all elements including magnetic ones and is particularly strong in heavy elements such as platinum and gold.

Our aim in this study is to prove the existence of the above monopole theoretically.
Since the spin-orbit interaction explicitly breaks the SU(2) invariance, we cannot follow the derivation of the HHM shown above.
Instead, we will directly calculate the effective electric and magnetic fields for the electron spin based on a nonequilibrium Green's function formalism, and derive Maxwell's equations they satisfy. 
 
We consider two types of spin-orbit interaction.
The first is that from a uniform field, ${\bm E}_{\rm R}$, namely the Rashba interaction~\cite{Rashba60}.
Such a field is realized at interfaces and surfaces~\cite{Meier07}.
The second is that from a random potential, $v_{\rm i}$, induced by heavy impurities.
The spin-orbit interaction thus reads
\begin{equation}
H_{\rm so}
=
-\frac{1}{\hbar}
( \lambda_{\rm R}{\bm E}_{\rm R}
-\lambda_{\rm i} {\bm \nabla} v_{\rm i} )
\cdot ({\bm p} \times {\bm \sigma}),
\end{equation}
where ${\bm p}$ is the electron momentum and $\lambda$ is the coupling constant (the subscripts $\rm R$ and $\rm i$ characterize Rashba and impurity-induced spin-orbit intereactions, respectively).
The interaction with the magnetization is described by 
 $H_{\rm sd}$.
The Hamiltonian of the present system is, therefore, given as $H = ({\bm p}^2 / 2m) +v_{\rm i} +H_{\rm sd} +H_{\rm so}$, where $m$ is the electron mass.
The magnetization we consider in $H_{\rm sd}$ is dynamic.
Dynamic magnetization, when coupled to the spin-orbit interaction, generates an electric charge flow~\cite{Hosono10,Takeuchi10,Hals10}.
The pumped electric current ${\bm j}$ is calculated by evaluating a quantum field theoretical expectation value of the electron velocity operator
$\hat{\bm v} = -(i\hbar / m) {\bm \nabla} +(1/\hbar) (\lambda_{\rm R} {\bm E}_{\rm R} -\lambda_{\rm i} {\bm \nabla} v_{\rm i}) \times {\bm \sigma}$.
By using field operators for electrons, $c^\dagger$ and $c$, the electric current thus reads
${\bm j} = -e {\rm tr} \langle{c^\dagger \hat{\bm v} c}\rangle$,
where ${\rm tr}$ denotes the trace over spin indices and the bracket represents the expectation value.
It is written in terms of the lesser component of the nonequilibrium Green's function~\cite{Haug07}, defined as
$G^<_{s s'}({\bm r},t;{\bm r}',t') \equiv (i/\hbar) \langle{c^\dagger_{s'}({\bm r}',t') c_s({\bm r},t)}\rangle$
($s$ and $s'$ are spin indices),
as
\begin{align}
{\bm j}({\bm r},t)
=
&e \, {\rm tr}
(
\{
\frac{\hbar^2}{2m} ({\bm \nabla}_{\bm r} -{\bm \nabla}_{{\bm r}'})
+i [\lambda_{\rm R} {\bm E}_{\rm R}
\notag
\\
&-\lambda_{\rm i} {\bm \nabla} v_{\rm i}({\bm r})] \times {\bm \sigma}
\}
G^<({\bm r},t;{\bm r}',t)
)_{{\bm r}'={\bm r}}.
\end{align}
This quantum field theoretical expectation value is evaluated by solving the Dyson's equation for the nonequilibrium Green's function defined on the Keldysh contour ($\rm C$),
\begin{align}
G_{s s'}({\bm r},t;{\bm r}',t')
=
&\delta_{s, s'}
g_s({\bm r},t;{\bm r}',t')
\notag
\\
&+\int{d^3r''} \int_{\rm C}{dt''}
g_s({\bm r},t;{\bm r}'',t'')
\notag
\\
&\times
(
\delta_{s, s''} v_{\rm i}({\bm r}'')
-J {\bm M}({\bm r}'',t'') \cdot {\bm \sigma}_{s s''}
\notag
\\
&+i \{
[\lambda_{\rm R} {\bm E}_{\rm R}
-\lambda_{\rm i} {\bm \nabla} v_{\rm i}({\bm r}'')]
\times {\bm \nabla}_{{\bm r}''}
\}
\cdot {\bm \sigma}_{s s''}
)
\notag
\\
&\times
G_{s'' s'}({\bm r}'',t'';{\bm r}',t'),
\end{align}
where $G_{s s'}({\bm r},t;{\bm r}',t') \equiv -(i/\hbar) \langle{{\rm T_C} [c_s({\bm r},t) c^\dagger_{s'}({\bm r}',t')]}\rangle$ 
(${\rm T_C}$ is the path-ordering operator)
and $g$ denotes the free Green's function.

\begin{figure}
\begin{center}
\includegraphics[scale=0.35]{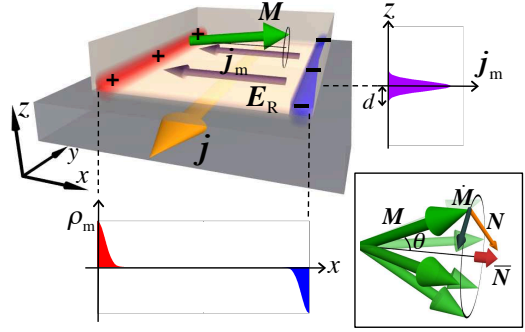}
\caption{
(Color online).
Schematic illustration of monopole pumping and detection in thin ferromagnetic film attached to a nonmagnetic layer.
Magnetization ($\bm M$) precession is induced by applying an oscillating magnetic field.
The Rashba field ${\bm E}_{\rm R}$ exists at the interface and creates the monopole current ${\bm j}_{\rm m}$ near the interface.
The width of the monopole current distribution, $d$,  is comparable to the decay length of the magnetization at the interface.
The monopole current induces the electric current ${\bm j}$ via Amp\`ere's law at the interface.
The impurity spin-orbit interaction directly induces positive ($+$) and negative ($-$) monopole charge distributions, $\rho_{\rm m}$, at the two edges.
This monopole distribution generates an electric current again perpendicular to the monopole current, as is seen from eqs.~(\ref{jresult}) - (\ref{AmperedivB}).
Inset: depiction of the spin damping vector ${\bm N}\equiv {\bm M} \times \dot{\bm M}$ arising from the precession of magnetization. 
The component of the spin damping vector perpendicular to the precession axis vanishes when time-averaged, leaving $\overline{\bm N}$ along the axis as the DC component.
}
\label{fig;system}
\end{center}
\end{figure}

In the calculation, the impurities are approximated as random point scatterers, and averaging is carried out as
$\langle{v_{\rm i}({\bm r}) v_{\rm i}({\bm r}')}\rangle_{\rm i} = n_{\rm i} u_{\rm i}^2 \delta^3({\bm r}-{\bm r}')$
($n_{\rm i}$ and $u_{\rm i}$ are the impurity concentration and the strength of scattering, respectively)~\cite{RammerQTT}.
The impurities give rise to an elastic lifetime for the electron, $\tau$, which is calculated in metals as 
$\tau = \hbar / 2\pi n_{\rm i}u_{\rm i}^2 \nu$ ($\nu$ is the density of states per volume)~\cite{lifetime}.
The Dyson's equation is solved by treating $\lambda$ and $J$ perturbatively to the first and second orders, respectively.
We consider a sufficiently slow dynamics of magnetization, namely, $\Omega \tau \ll 1$ ($\Omega$ is the frequency of magnetization dynamics), and assume that the magnetization structure varies smoothly in the space compared with the electron mean free path $\ell$, i.e., $q \ell \ll 1$ ($q$ is the wave number of magnetization profile).
The leading contribution in this case turns out to be~\cite{feynman,so}
\begin{align}
{\bm j}({\bm r},t)
=
&\frac{eJ^2}{V}
\sum_{{\bm k}, {\bm k}', {\bm q}_1,{\bm q}_2} \sum_{\omega, \Omega_1, \Omega_2}
e^{-i({\bm q}_1+{\bm q}_2)\cdot{\bm r} +i(\Omega_1+\Omega_2)t}
\Omega_1
\notag
\\
&\times
\frac{d f_\omega}{d \omega}
({\bm M}_{{\bm q}_1,\Omega_1} \times {\bm M}_{{\bm q}_2,\Omega_2})
\times
[
\frac{i \lambda_{\rm R} \tau}{\hbar}
{\bm E}_{\rm R}
|g^{\rm r}_{{\bm k},\omega}|^2
\notag
\\
&+\frac{4 \hbar^2 \lambda_{\rm i}}{3\pi \nu \tau^2}
({\bm q}_1 +{\bm q}_2)
\varepsilon_{\bm k}
|g^{\rm r}_{{\bm k},\omega}|^2
|g^{\rm r}_{{\bm k}',\omega}|^4
]
\notag
\\
&-D {\bm \nabla} \rho({\bm r},t),
\label{calculation}
\end{align}
where $V$ is the system volume,
$D \equiv 2 \varepsilon_{\rm F}\tau / 3m$ is the electron diffusion constant ($\varepsilon_{\rm F}$ represents the Fermi energy), $f_\omega$ is the Fermi distribution function, $g^{\rm r}$ is the retarded Green's function defined as $g^{\rm r}_{{\bm k},\omega} = [\hbar \omega -\varepsilon_{\bm k} +(i\hbar/2\tau)]^{-1}$, and $\varepsilon_{\bm k} = \hbar^2 {\bm k}^2 / 2m$.
The last term is the diffusive contribution arising from vertex corrections, where the electric charge density $\rho$ is~\cite{diffusion}
\begin{align}
\rho({\bm r},t)
=
&\frac{4e\nu \lambda_{\rm R} J^2 \tau^3}{\hbar^2 V}
{\bm \nabla} \cdot
\int{d^3r'} \int{dt'}
\notag
\\
&\times
\sum_{\bm q} \sum_\Omega
\frac{e^{-i {\bm q}\cdot({\bm r}-{\bm r}') +i \Omega(t-t')}}{D {\bm q}^2 \tau +i \Omega \tau}
\notag
\\
&\times
\{
{\bm E}_{\rm R} \times
[ {\bm M}({\bm r}',t') \times \dot{\bm M}({\bm r}',t') ]
\}.
\end{align}
Summing over the wave vectors and frequencies in eq.~(\ref{calculation}), the electric current is obtained as  
\begin{align}
{\bm j}
=
&-\frac{16e\nu \lambda_{\rm i} J^2 \varepsilon_{\rm F} \tau^2}{3\hbar^2}
{\bm \nabla} \times ({\bm M} \times \dot{\bm M})
\notag
\\
&-\frac{4e\nu \lambda_{\rm R} J^2 \tau^2}{\hbar^2}
{\bm E}_{\rm R} \times ({\bm M} \times \dot{\bm M})
-D {\bm \nabla} \rho.
\end{align}
This result is rewritten using the effective electric and magnetic fields, ${\bm E}_{\rm s}$ and ${\bm B}_{\rm s}$, respectively, as
($\mu$ and $\sigma_{\rm c}$ are the magnetic permeability and electric conductivity, respectively)
\begin{equation}
{\bm j} =
\frac{1}{\mu} {\bm \nabla} \times {\bm B}_{\rm s}
+\sigma_{\rm c} {\bm E}_{\rm s}
-D {\bm \nabla} \rho,
\label{jresult}
\end{equation}
where the effective fields are defined as~\cite{EMfield,arbitrariness}
\begin{align}
{\bm E}_{\rm s}
&\equiv
-\alpha_{\rm R}
{\bm E}_{\rm R} \times {\bm N},
\notag
\\
{\bm B}_{\rm s}
&\equiv
-\beta_{\rm i}
{\bm N}.
\label{EBdef}
\end{align}
Here, ${\bm N} \equiv {\bm M} \times \dot{\bm M}$ is a vector representing the spin damping torque (inset in Fig.~\ref{fig;system})~\cite{Chikazumi97}.
The coefficients $\alpha_{\rm R}$ and $\beta_{\rm i}$ are
$\alpha_{\rm R} \equiv 4e\nu \lambda_{\rm R} J^2 \tau^2 / \sigma_{\rm c} \hbar^2$
and
$\beta_{\rm i} \equiv 16e\nu \mu \lambda_{\rm i} J^2 \varepsilon_{\rm F} \tau^2 / 3\hbar^2$,
respectively.
The effective fields calculated here are those acting on the electronic spin in the same manner as the effective fields from the HHM.
Clearly, the fields [eq.~(\ref{EBdef})] do not satisfy Faraday's law or Gauss's law of conventional electromagnetism, but those with monopole contribution:
\begin{align}
{\bm \nabla} \times {\bm E}_{\rm s} +\dot{\bm B}_{\rm s} &= -{\bm j}_{\rm m}, \nonumber\\
{\bm \nabla} \cdot {\bm B}_{\rm s} &= \rho_{\rm m},
\label{AmperedivB}
\end{align}
where the monopole current and monopole density, respectively read
\begin{equation}
{\bm j}_{\rm m} =
\alpha_{\rm R} {\bm \nabla} \times ({\bm E}_{\rm R} \times {\bm N})
+\beta_{\rm i} \dot{\bm N},
\label{jmdef}
\end{equation}
and
\begin{equation}
\rho_{\rm m} =
-\beta_{\rm i} {\bm \nabla} \cdot {\bm N}.
\label{rhomdef}
\end{equation}

Equations~(\ref{AmperedivB}) - (\ref{rhomdef}) are the central results of this paper.
We have thus proved that a monopole exists when spin damping occurs, namely, a spin damping monopole.
The spin damping monopole is a composite object made from a magnetization configuration in the same manner as the HHM. 
It satisfies the conservation law $\dot{\rho}_{\rm m} +{\bm \nabla} \cdot {\bm j}_{\rm m} = 0$.

Our result obtained in a disordered system is a general one and can be extended to a clean system. 
In fact, the same monopole exists in a clean case, but only the coefficients $\alpha_{\rm R}$ and $\beta_{\rm i}$ appearing in the monopole density and current are changed. 
The same applies to the HHM; when we take into account the third-order contribution in $J$, our analysis correctly reproduces the HHM, which was discussed in a clean limit only.
Our approach is, therefore, a novel method of identifying monopoles.

The spin damping monopole is unique since it does not require a particular non-coplanar spin structure like a hedgehog, and so it exists quite generally in magnetic systems.
The simplest candidate for creating the monopole would be a thin ferromagnetic film put on a nonmagnetic material, as shown in Fig.~\ref{fig;system}.
We choose the $z$-axis perpendicular to the film.
A Rashba-type spin-orbit field would then arise at the interface along the $z$-direction~\cite{Meier07}.
We excite the precession of the uniform magnetization by applying the alternating magnetic field in the $yz$-plane in the presence of a static field along the $x$-axis (ferromagnetic resonance~\cite{Chikazumi97}).
The precession results in a spin damping vector with a finite time average, $\overline{\bm N}$, along $x$-direction (inset in Fig.~\ref{fig;system}).
In the present case with uniform magnetization, spatial derivatives in eqs.~(\ref{jmdef}) and (\ref{rhomdef}) arise at the interface and edges, where the magnetization vanishes.
The Rashba interaction contributes to the DC monopole current at the interface as
$\overline{j_{{\rm m}, x}^{\rm R}} = -\alpha_{\rm R} E_{\rm R} (\partial \overline{N} / \partial z) \simeq -(\alpha_{\rm R} / d) E_{\rm R} \overline{N}$,
where $d$ is the spatial scale of the magnetization decay at the interface.
The monopole current driven by random spin-orbit impurities, on the other hand, vanishes when time-averaged.
The total DC monopole current thus reads
$\overline{ {\bm j}_{\rm m} } = -{\bf e}_x (\alpha_{\rm R} / d) E_{\rm R} \overline{N}$
(${\bf e}_x$ represents the unit vector along the $x$-direction).
This monopole current at the interface generates an electromotive force along the $y$-direction via Amp\`ere's law for the monopole.
The monopole density induced by the random spin-orbit interaction arises at the edge of the ferromagnetic film since
${\bm \nabla} \cdot {\bm N} \simeq \partial {N_x} / \partial x$ is finite there.
The induced monopole density at the two edges is
$\overline{\rho_{\rm m}} = {\mp} (\beta_{\rm i} / d) \overline{N}$,
where the sign is positive on one side of the edge and negative on the other side.
The monopole then produces a magnetic field along the $x$-direction as
$\overline{{\bm B}_{\rm s}} = -{\bf e}_{x} \beta_{\rm i} \overline{N}$.
This field creates an electric current in the $y$-direction via the conventional Amp\`ere's law.
The total electric current density generated by the spin damping monopole [eq.~(\ref{jresult})] thus reduces to 
$\overline{\bm j} = -{\bf e}_y [\sigma_{\rm c} \alpha_{\rm R} E_{\rm R} +(\beta_{\rm i} / \mu d)] \overline{N}$.

When spin damping arises from the magnetization precession with a frequency $\Omega$ and an angle $\theta$, the monopole-induced current density is estimated as
$|\bar{j}| = (e k_{\rm F}^2 \Omega / \pi^2) (J\tau / \hbar)^2 \sin{\theta} [(\Delta_{\rm R} / \varepsilon_{\rm F}) +(4 / 3 k_{\rm F}d) (\Delta_{\rm i} / v_{\rm i})]$,
where $\Delta_{\rm R}$ and $\Delta_{\rm i}$ are the energy of the Rashba and impurity spin-orbit couplings, respectively
($k_{\rm F}$ is the Fermi wave vector).
In disordered ferromagnets, $J / \varepsilon_{\rm F} \sim 0.1$, $\varepsilon_{\rm F}\tau / \hbar \sim 10$ and $k_{\rm F}^{-1} \sim 2$ \AA.
The Rashba interaction can be enhanced on surfaces and at interfaces, resulting in $\Delta_{\rm R} / \varepsilon_{\rm F} \sim 0.1$ ($\Delta_{\rm i} / \varepsilon_{\rm F}$ is generally smaller)~\cite{Ast07}.
When $\theta = 30^\circ$ and $\Omega = 1$ GHz, the electric current density is thus $2 \times 10^7$ A/m$^2$ which is sufficiently large for experimental detection.
In addition to DC, there is an AC component in eq.~(\ref{jresult}), which would be accessible by time-resolved measurement.

We note that the electric current estimated here is an initial current that arises when the pumping of monopoles starts. 
When the monopole current is pumped steadily, monopole accumulation grows at the edges of the system, inducing a diffusive current.
The steady monopole distribution is then determined by the balance of this backward diffusion and the pumped monopole current.

Direct evidence of the spin damping monopole is given by detecting the electric current discussed above.
Surprisingly, the signal from the spin damping monopole might have already been detected.
In fact, the electric voltage due to magnetization precession has been observed in a junction of a ferromagnet on a Pt film in a pioneering work by Saitoh {\it et al.}~\cite{Saitoh06}.
The mechanism of voltage generation has been explained by the inverse spin Hall effect. 
According to the inverse spin Hall scenario, magnetization precession generates a spin current via the spin pumping effect~\cite{Tserkovnyak02}, and the spin current ${\bm j}_{\rm s}$ is converted into an electric current by the spin-orbit interaction (the inverse of the spin Hall effect).
This explanation assumes that 
the conversion mechanism of 
$j_i = \epsilon_{ijk} j_{{\rm s}, j}^k$
where $k$ is the index representing the spin polarization of the spin current~\cite{Hirsch99}.
A recent theoretical study has revealed, however, that the conversion formula is not universal; it does not apply to the slowly varying magnetization configuration or in the presence of disorder~\cite{Takeuchi10}.
Rather, the conversion formula is an approximated one connecting a physical observable (electric current) to a spin current whose definition depends on the specific system considered. 
Our result obtained in the present paper suggests a different scenario that is universal owing to the symmetry of Maxwell's equations.

For the experimental confirmation of the spin damping monopole, of crucial importance is the separation of the monopole signal from the inverse spin Hall signal driven by a spin current.
This is accomplished by applying an electric field (${\bm E}_{\rm s}$) perpendicular to the junction of Fig.~\ref{fig;system}.
The monopole contribution then leads to a transverse electric current as a result of the Hall effect of the monopole~\cite{Lorentz}, while the contribution of the spin current is not affected.
In another experiment, the strongest evidence of the spin damping monopole is given by observing a magnetic field (${\bm B}_{\rm s}$) produced by monopoles via Gauss's law by electron holography.

We have shown analytically that a magnetic monopole is a common object in dynamic magnetic systems with damping. 
The control of spin damping monopoles is as feasible as that of electrons, for both are governed symmetrically by Maxwell's equations of electromagnetism.
We here propose the monopolotronics, i.e., the control of monopoles, as a novel concept of realizing spintronic devices.


\begin{acknowledgements}
The authors are grateful to N. Kitazawa, R. Matsumoto, S. Murakami, and N. Nagaosa for discussion.
This work was supported by 
a Grant-in-Aid for Scientific Research in Priority Areas (Grant No. 1948027) from the Ministry of Education, Culture, Sports, Science and Technology,
a Grant-in-Aid for Scientific Research (B) (Grant No. 22340104) from Japan Society for the Promotion of Science.
One of the authors (A.T.) is financially supported by the Japan Society for the Promotion of Science for Young Scientists.
\end{acknowledgements}



\begin{thebibliography}{99}
\bibitem{lifetime}
Small corrections on the order of $\hbar / \varepsilon_{\rm F}\tau$ are neglected~\cite{RammerQTT}.
This corresponds to the Born approximation.
\bibitem{feynman}
Equation~(\ref{calculation}) is obtained by evaluating three and six distinct self-energy types of Feynman diagrams for the the Rashba and impurity-induced spin-orbit interactions, respectively, and vertex corrections including infinite summation of the impurity scattering~\cite{RammerQTT}.
\bibitem{EMfield}
In eq.~(\ref{jresult}), electron diffusion ($D {\bm \nabla} \rho$) induces the effective electric and magnetic polarizations ${\bm P}_{\rm s}$ and ${\bm M}_{\rm s}$, respectively.
By using the relation $(-D{\bm \nabla}^2 +\partial_t)\rho = -\sigma_{\rm c} {\bm \nabla} \cdot {\bm E}_{\rm s}$,
the electric current [eq.~(\ref{jresult})] is described as the rotation of the magnetic field and the time derivative of the electric field, ${\bm j} = {\bm \nabla} \times {\bm H}_{\rm s} -\dot{\bm D}_{\rm s}$, where ${\bm H}_{\rm s}$ and ${\bm D}_{\rm s}$ are the fields defined as ${\bm H}_{\rm s} \equiv (1/\mu) {\bm B}_{\rm s} -{\bm M}_{\rm s}$ and ${\bm D}_{\rm s} \equiv \varepsilon {\bm E}_{\rm s} +{\bm P}_{\rm s}$ ($\varepsilon = -\sigma_{\rm c} \tau$ being the permittivity), respectively.
Thus, ${\bm E}_{\rm s}$ and ${\bm B}_{\rm s}$ are identified with the electric and magnetic fields, respectively.
\bibitem{arbitrariness}
Equations~(\ref{jresult}) and (\ref{EBdef}) apparently contain an arbitrariness, ${\bm E}_{\rm s} \rightarrow {\bm E}_{\rm s} +(1/\sigma_{\rm c}) {\bm \nabla} \times {\bm C}$ and ${\bm B}_{\rm s} \rightarrow {\bm B}_{\rm s} -\mu {\bm C}$, where ${\bm C}$ is an arbitrary vector field.
However, such a transform in Maxwell's equations with monopoles is generally not allowed because of the gauge invariance in the original space with a higher symmetry.
In fact, in the case of HHM, the SU(2) gauge invariance forbids this arbitrariness~\cite{Volovik87}.
We expect the same argument applies to the present spin damping monopole.
\end{thebibliography}
\end{document}